\documentclass[english]{article}
\usepackage[T1]{fontenc}
\usepackage[latin9]{inputenc}
\usepackage{color}
\usepackage{graphicx}
\usepackage{esint}
\usepackage{babel}
\begin{document}

\title{\textbf{New proof of general relativity through the correct physical
interpretation of the Mössbauer rotor experiment}}

\author{\textbf{Christian Corda}}
\maketitle
\begin{center}
Research Institute for Astronomy and Astrophysics of Maragha (RIAAM),
P.O. Box 55134-441, Maragha, Iran and International Institute for
Applicable Mathematics \& Information Sciences (IIAMIS),  B.M. Birla
Science Centre, Adarsh Nagar, Hyderabad - 500 463, India 
\par\end{center}

\begin{center}
\textit{E-mail address:} \textcolor{blue}{cordac.galilei@gmail.com}
\par\end{center}
\begin{abstract}
In this Essay, we give a correct interpretation of a historical experiment
by Kündig on the transverse Doppler shift in a rotating system (Mössbauer
rotor experiment). This experiment has been recently first reanalyzed,
and then replied by an experimental research group. The results of
reanalyzing the experiment have shown that a correct re-processing
of Kündig's experimental data gives an interesting deviation of a
relative redshift between emission and absorption resonant lines from
the standard prediction based on the relativistic dilatation of time.
Subsequent new experimental results by the reply of Kündig experiment
have shown a deviation from the standard prediction even higher. By
using the Equivalence Principle (EP), which states the equivalence
between the gravitational \textquotedbl{}force\textquotedbl{} and
the \emph{pseudo-force} experienced by an observer in a non-inertial
frame of reference (included a rotating frame of reference), here
the theoretical framework of the Mössbauer rotor experiment is reanalyzed
directly in the rotating frame of reference through a\emph{ }general
relativistic treatment. It will be shown that previous analyses missed
an important effect of clock synchronization. By adding this new effect,
the correct general relativistic prevision is in perfect agreement
with the new experimental results. Such an effect of clock synchronization
has been missed in various papers in the literature, with some subsequent
claim of invalidity of the relativity theory and/or some attempts
to explain the experimental results through ``exotic'' effects.
The general relativistic interpretation in this Essay shows, instead,
that the new experimental results of the Mössbauer rotor experiment
are a new, strong and independent, proof of general relativity.
\end{abstract}
\emph{Essay written for the Gravity Research Foundation 2018 Awards
for Essays on Gravitation: Honorable Mention Winner.}
\begin{quotation}
\medskip{}

\textbf{To the memory of Enrico Lista.}
\end{quotation}
We give a correct interpretation of a historical experiment by Kündig
on the transverse Doppler shift in a rotating system, measured with
the Mössbauer effect (Mössbauer rotor experiment) \cite{key-3}. The
Mössbauer effect (discovered by R. Mössbauer in 1958 \cite{key-14})
consists in resonant and recoil-free emission and absorption of gamma
rays, without loss of energy, by atomic nuclei bound in a solid. It
resulted and currently results very important for basic research in
physics and chemistry. In this Essay, we will focus on the so called
Mössbauer rotor experiment. In this particular experiment, the Mössbauer
effect works through an absorber orbited around a source of resonant
radiation (or vice versa). The aim is to verify the relativistic time
dilation for a moving resonant absorber (the source), inducing a relative
energy shift between emission and absorption lines. 

In a couple of recent papers \cite{key-1,key-2}, the authors first
reanalyzed in \cite{key-1} the data of a known experiment of Kündig
on the transverse Doppler shift in a rotating system, measured with
the Mössbauer effect \cite{key-3}, and second, they carried out their
own experiment on the time dilation effect in a rotating system \cite{key-2}.
In \cite{key-1}, it has been found that the original experiment by
Kündig \cite{key-3} contained errors in the data processing. A puzzling
fact was that, after correction of the errors of Kündig, the experimental
data gave the value \cite{key-1}

\begin{equation}
\frac{\nabla E}{E}\simeq-k\frac{v^{2}}{c^{2}},\label{eq: k}
\end{equation}
where $k=0.596\pm0.006$, instead of the standard relativistic prediction
$k=0.5$ due to time dilatation. The authors of \cite{key-1} stressed
that the deviation of the coefficient $k$ in equation (\ref{eq: k})
from $0.5$ exceeds by almost 20 times the measuring error and that
the revealed deviation cannot be attributed to the influence of rotor
vibrations and other disturbing factors. All these potential disturbing
factors have been indeed excluded by a perfect methodological trick
applied by Kündig \cite{key-3}, that is a first-order Doppler modulation
of the energy of $\gamma-$quanta on a rotor at each fixed rotation
frequency. In that way, Kündig's experiment can be considered as the
most precise among other experiments of the same kind {[}4\textendash 8{]},
where the experimenters measured only the count rate of detected $\gamma-$quanta
as a function of rotation frequency. The authors of \cite{key-1}
have also shown that the experiment {[}8{]}, which contains much more
data than the ones in {[}4\textendash 7{]}, also confirms the supposition
$k>0.5.$ Motivated by their results in \cite{key-1}, the authors
carried out their own experiment \cite{key-2}. They decided to repeat
neither the scheme of the Kündig experiment \cite{key-3}, nor the
schemes of other known experiments on the subject previously mentioned
above {[}4\textendash 8{]}. In that way, they got independent information
on the value of $k$ in equation (\ref{eq: k}). In particular, they
refrained from the first-order Doppler modulation of the energy of
$\gamma-$quanta, in order to exclude the uncertainties in the realization
of this method \cite{key-2}. They followed the standard scheme {[}4\textendash 8{]},
where the count rate of detected $\gamma-$quanta $N$ as a function
of the rotation frequency $\nu$ is measured. On the other hand, differently
from the experiments {[}4\textendash 8{]}, they evaluated the influence
of chaotic vibrations on the measured value of $k$ \cite{key-2}.
Their developed method involved a joint processing of the data collected
for two selected resonant absorbers with the specified difference
of resonant line positions in the Mössbauer spectra \cite{key-2}.
The result obtained in \cite{key-2} is $k=0.68\pm0.03$, confirming
that the coefficient $k$ in Eq. (\ref{eq: k}) substantially exceeds
0.5. The scheme of the new Mössbauer rotor experiment is in Figure
1, while technical details on it can be found in \cite{key-2}. 

\begin{figure}
\includegraphics[scale=0.75]{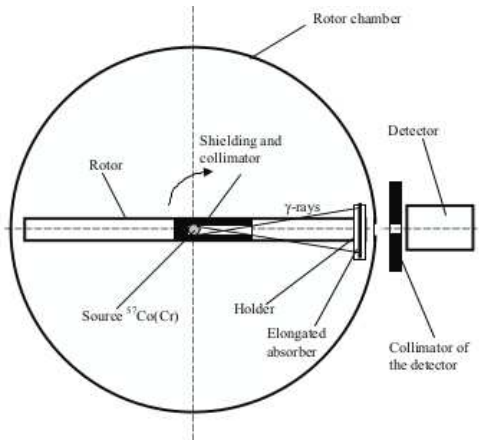}

\caption{Scheme of the new Mössbauer rotor experiment, adapted from ref. \cite{key-2}}
\end{figure}

In this Essay, the EP, which states the equivalence between the gravitational
\textquotedbl{}force\textquotedbl{} and the \emph{pseudo-force} experienced
by an observer in a non-inertial frame of reference (included a rotating
frame of reference), will be used to reanalyze the theoretical framework
of Mössbauer rotor experiments directly in the rotating frame of reference,
by using a full general relativistic treatment \cite{key-16}. The
results will show that previous analyses missed an important effect
of clock synchronization and that the correct general relativistic
prevision gives $k\simeq\frac{2}{3}$ \cite{key-16}, in perfect agreement
with the new experimental results of \cite{key-2}. In that way, the
general relativistic interpretation of this Essay shows that the new
experimental results of the Mössbauer rotor experiment are a new,
strong and independent, proof of general relativity. We also stress
that various papers in the literature (included ref. \cite{key-4}
published in Phys. Rev. Lett.) missed the effect of clock synchronization
{[}1\textendash 8{]}, {[}11\textendash 13{]} with some subsequent
claim of invalidity of relativity theory and/or some attempts to explain
the experimental results through ``exotic'' effects \cite{key-1,key-2,key-11,key-12,key-13}.

Following \cite{key-9,key-16}, one considers a transformation from
an inertial frame, in which the space-time is Minkowskian, to a rotating
frame of reference. Using cylindrical coordinates, the line element
in the starting inertial frame is \cite{key-9,key-16}

\begin{equation}
ds^{2}=c^{2}dt^{2}-dr^{2}-r^{2}d\phi^{2}-dz^{2}.\label{eq: Minkowskian}
\end{equation}
The transformation to a frame of reference $\left\{ t',r',\phi'z'\right\} $
rotating at the uniform angular rate $\omega$ with respect to the
starting inertial frame is given by \cite{key-9,key-16}

\begin{equation}
\begin{array}{cccc}
t=t'\; & r=r' & \;\phi=\phi'+\omega t'\quad & z=z'\end{array}.\label{eq: trasformazione Langevin}
\end{equation}
Thus, Eq. (\ref{eq: Minkowskian}) becomes the following well-known
line element (Langevin metric) in the rotating frame \cite{key-9,key-16}
\begin{equation}
ds^{2}=\left(1-\frac{r'^{2}\omega^{2}}{c^{2}}\right)c^{2}dt'^{2}-2\omega r'^{2}d\phi'dt'-dr'^{2}-r'^{2}d\phi'^{2}-dz'^{2}.\label{eq: Langevin metric}
\end{equation}
The transformation (\ref{eq: trasformazione Langevin}) is both simple
to grasp and highly illustrative of the general covariance of general
relativity as it shows that one can work first in a \textquotedbl{}simpler\textquotedbl{}
frame and then transforming to a more \textquotedbl{}complex\textquotedbl{}
one \cite{key-16}. As one considers light propagating in the radial
direction ($d\phi'=dz'=0$), the line element (\ref{eq: Langevin metric})
reduces to \cite{key-16}

\begin{equation}
ds^{2}=\left(1-\frac{r'^{2}\omega^{2}}{c^{2}}\right)c^{2}dt'^{2}-dr'^{2}.\label{eq: metrica rotante}
\end{equation}
The EP permits to interpret the line element (\ref{eq: metrica rotante})
in terms of a curved spacetime in presence of a static gravitational
field \cite{key-10,key-15,key-16}. In that way, one obtains a purely
general relativistic interpretation of the pseudo-force experienced
by an observer in a rotating, non-inertial frame of reference \cite{key-16}.
Setting the origin of the rotating frame in the source of the emitting
radiation, one gets a first contribution, which arises from the ``gravitational
redshift'', that can be directly computed using Eq. (25.26) in \cite{key-10},
which, in the twentieth printing 1997 of \cite{key-10}, is written
as 
\begin{equation}
z\equiv\frac{\Delta\lambda}{\lambda}=\frac{\lambda_{received}-\lambda_{emitted}}{\lambda_{emitted}}=|g_{00}(r'_{1})|^{-\frac{1}{2}}-1\label{eq: z  MTW}
\end{equation}
and represents the redshift of a photon emitted by an atom at rest
in a gravitational field and received by an observer at rest at infinity.
Here, a slightly different equation with respect to Eq. (25.26) in
\cite{key-10} will be used, because here one considers a gravitational
field which increases with increasing radial coordinate $r'$, while
Eq. (25.26) in \cite{key-10} concerns a gravitational field which
decreases with increasing radial coordinate \cite{key-16}. Also,
the zero potential is set in $r'=0$ instead of at infinity, and one
uses the proper time $\tau$ instead of the wavelength $\lambda$
\cite{key-16}. Thus, by using Eq. (\ref{eq: metrica rotante}), one
gets \cite{key-16} 
\begin{equation}
\begin{array}{c}
z_{1}\equiv\frac{\nabla\tau_{10}-\nabla\tau_{11}}{\tau}=1-|g_{00}(r'_{1})|{}^{-\frac{1}{2}}=1-\frac{1}{\sqrt{1-\frac{\left(r'_{1}\right)^{2}\omega^{2}}{c^{2}}}}\\
\\
=1-\frac{1}{\sqrt{1-\frac{v^{2}}{c^{2}}}}\simeq-\frac{1}{2}\frac{v^{2}}{c^{2}},
\end{array}\label{eq: gravitational redshift}
\end{equation}
where $\nabla\tau_{10}$ is the delay of the emitted radiation, $\nabla\tau_{11}$
is the delay of the received radiation, $r'_{1}\simeq c\tau$ is the
radial distance between the source and the detector and $v=r'_{1}\omega$
is the tangential velocity of the detector \cite{key-16}. Hence,
one finds a first contribution, say $k_{1}=\frac{1}{2}$, to $k$
\cite{key-16}. We stress again that the power of the EP enabled us
to use a pure general relativistic treatment in the above discussion
\cite{key-16}.

Now, one notices that the variations of proper time $\nabla\tau_{10}$
and $\nabla\tau_{11}$ have been calculated in the origin of the rotating
frame which is located in the source of the radiation \cite{key-16}.
But the detector is moving with respect to the origin in the rotating
frame \cite{key-16}. Thus, the clock in the detector must be synchronized
with the clock in the origin, and this gives a second, additional,
contribution to the redshift \cite{key-16}, which was missed in previous
analyses {[}1\textendash 8{]}, {[}11\textendash 13{]}. To compute
this second contribution, one uses Eq. (10) of \cite{key-9}, which
represents the proper time increment $d\tau$ on the moving clock
having radial coordinate $r'$ for values $v\ll c$ 

\begin{equation}
d\tau=dt'\left(1-\frac{r'^{2}\omega^{2}}{c^{2}}\right).\label{eq:secondo contributo}
\end{equation}
Inserting the condition of null geodesics $ds=0$ in Eq. (\ref{eq: metrica rotante}),
one gets \cite{key-16} 
\begin{equation}
cdt'=\frac{dr'}{\sqrt{1-\frac{r'^{2}\omega^{2}}{c^{2}}}},\label{eq: tempo 2}
\end{equation}
where the positive sign in the square root has been taken, because
the radiation is propagating in the positive $r$ direction \cite{key-16}.
Combining eqs. (\ref{eq:secondo contributo}) and (\ref{eq: tempo 2}),
one obtains \cite{key-16}

\begin{equation}
cd\tau=\sqrt{1-\frac{r'^{2}\omega^{2}}{c^{2}}}dr'.\label{eq: secondo contributo finale}
\end{equation}
Eq. (\ref{eq: secondo contributo finale}) is well approximated by
\cite{key-16} 
\begin{equation}
cd\tau\simeq\left(1-\frac{1}{2}\frac{r'^{2}\omega^{2}}{c^{2}}+....\right)dr',\label{eq: well approximated}
\end{equation}
which permits to find the second contribution of order $\frac{v^{2}}{c^{2}}$
to the variation of proper time as \cite{key-16} 
\begin{equation}
c\nabla\tau_{2}=\int_{0}^{r'_{1}}\left(1-\frac{1}{2}\frac{\left(r'_{1}\right)^{2}\omega^{2}}{c^{2}}\right)dr'-r'_{1}=-\frac{1}{6}\frac{\left(r'_{1}\right)^{3}\omega^{2}}{c^{2}}=-\frac{1}{6}r'_{1}\frac{v^{2}}{c^{2}}.\label{eq: delta tau 2}
\end{equation}
Thus, as $r'_{1}\simeq c\tau$ is the radial distance between the
source and the detector, one gets the second contribution of order
$\frac{v^{2}}{c^{2}}$ to the redshift as \cite{key-16} 
\begin{equation}
z_{2}\equiv\frac{\nabla\tau_{2}}{\tau}=-k_{2}\frac{v}{c^{2}}^{2}=-\frac{1}{6}\frac{v^{2}}{c^{2}}.\label{eq: z2}
\end{equation}
Then, one obtains $k_{2}=\frac{1}{6}$ and, using eqs. (\ref{eq: gravitational redshift})
and (\ref{eq: z2}), the total redshift is \cite{key-16} 
\begin{equation}
\begin{array}{c}
z\equiv z_{1}+z_{2}=\frac{\nabla\tau_{10}-\nabla\tau_{11}+\nabla\tau_{2}}{\tau}=-\left(k_{1}+k_{2}\right)\frac{v^{2}}{c^{2}}\\
\\
=-\left(\frac{1}{2}+\frac{1}{6}\right)\frac{v^{2}}{c^{2}}=-k\frac{v^{2}}{c^{2}}=-\frac{2}{3}\frac{v^{2}}{c^{2}}=0.\bar{6}\frac{v^{2}}{c^{2}},
\end{array}\label{eq: z totale}
\end{equation}
which is completely consistent with the result $k=0.68\pm0.03$ in
\cite{key-2}.

We stress that the additional factor $-\frac{1}{6}$ in Eq. (\ref{eq: z2})
comes from clock synchronization \cite{key-16}. In other words, its
theoretical absence in the works {[}1\textendash 8{]}, {[}11\textendash 13{]}
reflected the incorrect comparison of clock rates between a clock
at the origin and one at the detector \cite{key-16}. This generated
wrong claims of invalidity of relativity theory and/or some attempts
to explain the experimental results through ``exotic'' effects \cite{key-1,key-2,key-11,key-12,key-13}
which, instead, must be rejected. Notice that, even in discussing
the effect of clock synchronization, a pure general relativistic treatment
has been performed.

The appropriate reference \cite{key-9} has been evoked for a discussion
of the Langevin metric. This is dedicated to the use of general relativity
in Global Positioning Systems (GPS), which leads to the following
interesting realization \cite{key-16}: the correction of $-\frac{1}{6}$
in Eq. (\ref{eq: z2}) is analogous to the correction that one must
consider in GPS when accounting for the difference between the time
measured in a frame co-rotating with the Earth geoid and the time
measured in a non-rotating (locally inertial) Earth centered frame
(and also the difference between the proper time of an observer at
the surface of the Earth and at infinity). Indeed, if one simply considers
the gravitational redshift due to the Earth's gravitational field,
but neglects the effect of the Earth's rotation, GPS would not work
\cite{key-16}! The key point is that the proper time elapsing on
the orbiting GPS clocks cannot be simply used to transfer time from
one transmission event to another because path-dependent effects must
be taken into due account, exactly like in the above discussion of
clock synchronization \cite{key-16}. In other words, the obtained
correction $-\frac{1}{6}$ in Eq. (\ref{eq: z2}) is not an obscure
mathematical or physical detail, but a fundamental ingredient that
must be taken into due account \cite{key-16}. Further details on
the analogy between the results of this Essay and the use of general
relativity in GPS have been highlight in \cite{key-16}. 

\subsection*{Conclusion remarks}

In this Essay, the power of the EP, which states the equivalence between
the gravitational \textquotedbl{}force\textquotedbl{} and the \emph{pseudo-force}
experienced by an observer in a non-inertial frame of reference (included
a rotating frame of reference), has been used to reanalyze, from a
pure general relativistic point of view, the theoretical framework
of the new Mössbauer rotor experiment in \cite{key-2}, directly in
the rotating frame of reference. The results have shown that previous
analyses missed an important effect of clock synchronization and that
the correct general relativistic prevision gives $k\simeq\frac{2}{3}$,
in perfect agreement with the new experimental results in \cite{key-2}.
Thus, in this Essay it has been shown that the general relativistic
interpretation of the new experimental results of the Mössbauer rotor
experiment is a new, strong and independent, proof of Einstein general
relativity. The importance of the results in this Essay is stressed
by the issue that various papers in the literature (included ref.
\cite{key-4} published in Phys. Rev. Lett.) missed the effect of
clock synchronization {[}1\textendash 8{]}, {[}11\textendash 13{]},
with some subsequent claim of invalidity of relativity theory and/or
some attempts to explain the experimental results through ``exotic''
effects \cite{key-1,key-2,key-11,key-12,key-13}, which, instead,
must be rejected. 

\subsection*{Acknowledgements}

This Essay has been supported financially by the Research Institute
for Astronomy and Astrophysics of Maragha (RIAAM).

\end{document}